\documentclass[journal,twoside,web]{ieeecolor}

\let\labelindent\relax

\usepackage{generic}
\usepackage{cite}
\usepackage{amsmath,amssymb,amsfonts}
\usepackage{algorithmic}
\usepackage{graphicx}
\usepackage{textcomp}
\usepackage{comment}
\usepackage{stmaryrd}
\usepackage{caption}
\usepackage{float} 
\usepackage{enumitem} 

\usepackage{algorithmic,algorithm}
\usepackage{xcolor}
\definecolor{blue}{RGB}{51,131,200}

\def\BibTeX{{\rm B\kern-.05em{\sc i\kern-.025em b}\kern-.08em
    T\kern-.1667em\lower.7ex\hbox{E}\kern-.125emX}}

\begin{document}

\title{Continuous emission ultrasound: a new paradigm to ultrafast ultrasound imaging}
\author{Axel Adam, Barbara Nicolas, Adrian Basarab, \IEEEmembership{Senior Member, IEEE}, and Hervé Liebgott, \IEEEmembership{Member, IEEE}
\thanks{Copyright (c) 2021 IEEE. Personal use of this material is permitted. However, permission to use this material for any other purposes must be obtained from the IEEE by sending an email to pubs-permissions@ieee.org.}}

\maketitle

\begin{abstract}
Current imaging techniques in echography rely on the pulse-echo (PE) paradigm which provides a straight-forward access to the in-depth structure of tissues. They inherently face two major challenges: the limitation of the pulse repetition frequency, directly linked to the imaging framerate, and, due to the emission scheme, their blindness to the phenomena that happen in the medium during the majority of the acquisition time. To overcome these limitations, we propose a new paradigm for ultrasound imaging, denoted by continuous emission ultrasound imaging (CEUI) \cite{CEUIpatent2023}, for a single input single output (SISO) device.

A continuous insonification of the medium is done by the probe using a coded waveform inspired from the radar and sonar literature. A framework coupling a sliding window approach (SWA) and pulse compression methods processes the recorded echoes to rebuild a motion-mode (M-mode) image from the medium with a high temporal resolution compared to state-of-the-art ultrafast imaging methods. 

A study on realistic simulated data, with regards to the motion of the medium, has been carried out and, achieved results assess an unequivocal improvement of the slow time frequency up to, at least, two orders of magnitude compared to ultrafast US imaging methods. This enhancement leads, therefore, to a ten times improvement in the temporal separability of the imaging system.  
In addition, it demonstrates the capability of CEUI to catch relatively short and quick events, in comparison to the imaging period of PE methods, at any instant of the acquisition. 


\end{abstract}


\section{Introduction}

%
Echography, also known as ultrasound (US) imaging, is a non-invasive medical imaging technique primarily used to characterize the cross-section of local anatomy by estimating the reflectivity map of tissues. 
%
%
In recent decades, there has been significant interest in advancing towards three-dimensional (3D) US imaging as a diagnosis tool in numerous applications \cite{huang_review_2017}. 
%
However, this new technology is facing various challenges, among which, the use of huge amount of piezo-electric elements. This problem was addressed using row-column (RC) arrays \cite{jensen_anatomic_2022}, sparse arrays or compound lens for instance \cite{engholm2018increasing}.
%
Second, the constrained imaging framerate poses a substantial impediment to the progress of 3D US imaging as suggested in this review \cite{huang_review_2017}.
Efforts to overcome this challenge and enhance the capabilities of 3D US imaging systems are essential for improving diagnostic accuracy and expanding the scope of medical applications \cite{provost_3d_2014}. 

%
More generally, diagnosis methods of cardiovascular pathologies must deal with highly dynamic medium.
For instance, abdominal aortic aneurysm diagnosis using Pulse-Wave Velocity (PWV) deals with a several meters per second wave propagating during few milliseconds \cite{labovitz1986quantitative}.
In addition, 4D ultrasound flow imaging faces difficulties when dealing with high flow peak velocity cases, up to 1 $m.s^{-1}$, exceeding the Nyquist velocity limit of the imaging system \cite{correia20164d}. 
Hence, there is a significant need for imaging systems with sufficient temporal resolution to accurately depict the dynamics of the medium.


%

With the aim of capturing faster and shorter events by increasing the imaging framerate, significant innovations have been made, especially in two-dimensional (2D) US imaging. The latter rely on a  specific use of the probe elements at emission such as synthetic transmit aperture (STA) documented in \cite{jensen_synthetic_2006} or multiline transmission (MLT) discussed in \cite{badescu2019comparison} and compared to diverging wave imaging (DWI). The latter approach relies on the design of the emission of a specific cylindrical front wave shape, similarly to coherent plane wave compounding \cite{montaldo_coherent_2009} which is based on a multiple insonification by pulses with a plane wavefront in different propagation angles. Coded excitation techniques, as presented in \cite{misaridis_use_2005}, enable to outperform the framerate of conventional US imaging systems. A mix of a specific emission scheme with coded emissions can be used to boost the temporal resolution \cite{gran_spatial_2008}. 

Despite these innovations, it is important to note that all these techniques still operate within the pulse-echo (PE) paradigm: 
a subset of piezoelectric elements is excited within an US probe to emit a relatively brief pulse through the medium under examination. 
Subsequently, the US waves interact with the echogenic tissues, leading to backscattering.
The reflected signals are recorded by the receiving elements of the probe, contributing to the generation of images and the extraction of specific features.
However, this paradigm imposes two challenging limitations, the foremost being the constraint on imaging framerate, while the second being the missing interaction in the medium.

The limitation of the imaging framerate is due to an upper bound set on the pulse repetition frequency (PRF) to ensure the proper running of these techniques.  
The reflectivity map reconstruction of PE-based imaging methods, like delay-and-sum (DAS) \cite{perrot_so_2021}, relies mostly on the time-of-flight (TOF) between the instant of pulse emission and the reception instants of the resulting echoes. Crossing the information contained by the recorded radio-frequency (RF) signals by all the receivers allows for the estimation of a 2D or 3D representation of the medium.
%
%
Spatial ambiguity at reception occurs when a specific receiver element of the probe simultaneously receives echoes from tissues within the insonified field of view at distinct distances from the given sensor. In this situation, echogenic tissues located at different ranges from the receiver generate echoes corresponding to different pulses emitted by the probe: the false imaged structures are called range ambiguity artifacts (RAA) \cite{ng_resolution_2011}. the TOF of echoes becomes ambiguous as the pulse that generates each respective echo is no longer known. To prevent from this issue, the probe elements must refrain from emitting a new pulse to capture the subsequent representation of the medium until all echoes produced by the previous emitted pulse are received. Therefore, this constrains the PRF depending on the desired maximal imaging range.

The second limitation stems from the brief duration of emitted pulses. In the pursuit of imaging rapid and transient events in echography, PE is therefore, inherently challenged. The interaction time between the emission and the tissues is significantly short in comparison to the pulse repetition interval.
%
As a result, the observation of the medium occurs over a limited time frame, resulting in a lack of echoes that carry the information necessary for reconstructing a reflectivity map during a long period of time. Even if coded excitation increase the length of the emission compared to a conventional monochromatic pulse, it is imperative to keep the pulse temporal width as short as possible to ensure good imaging performances.
While pulse compression on RF signals can address the issue of a deterioration of the axial resolution if a longer pulse is sent, it still depends largely on probe bandwidth (BW) \cite{rao1994investigation}. 
Moreover, this constraint is also dictated by hardware limitations: when probe elements serve as emitters, they cannot simultaneously function as receivers, preventing the imaging of the near field as discussed in \cite{tiran_multiplane_2015}. 

Given the two aforementioned limitations intrinsic to PE-based imaging techniques, our paper introduces a novel paradigm for US imaging. Operating under the assumption that an US probe with such capabilities or an equivalent setup of two probes exist, we propose the continuous emission of US signals into the medium \cite{CEUIpatent2023} as similarly investigated in radar \cite{blunt2016overview} and sonar \cite{munafo_continuous_2019} \cite{yin_integrated_2020} \cite{hickman_non-recurrent_2012}  applications.
This emission strategy produces echoes from every region within the field of view throughout the entire acquisition period. The allotted receiving elements then record a weighted summation of backscattered signals from potentially multiple regions at different distances. Consequently, the continuous emission approach ensures that the probe constantly monitors the entire medium, capturing even rapid and fleeting events that PE-based methods struggle to observe. This is achieved by creating an ongoing interaction with the entire field of view at any given moment during the acquisition.

Toward the goal to develop a continuous emission ultrasound imaging (CEUI) system, this work achieves the next objective: the design of a first framework for CEUI employing two mono-element probes, currently implemented through simulation.
Indeed, the prevailing design of current ultrasonic systems mandates the piezoelectric elements to serve as receivers the majority of the acquisition duration to prevent from overheating. 
Acknowledging this constraint, our research is strategically focused on formulating a CEUI system designed to yield a highly temporally resolved one-dimensional (1D) monitoring of the medium in the form of a motion-mode (M-Mode) image. A single input single output (SISO) device is considered using two distinct mono-element probes. The slow time resolution of the M-Mode will be two orders of magnitude higher compared to ultrafast PE approaches. 
Because current simulators such as Field-II \cite{jensen_calculation_1992}, \cite{jensen_field_1996}, MUST \cite{garcia_make_2021} or k-Wave \cite{treeby_k_wave_2010} are not capable to model realistically non-stationary media, we will propose a model to generate simulated US data generated by such a medium. Indeed, using the aforementioned simulators, one must assume a multi-static medium as they can not model a moving medium while it is interacting  with emitted waveform.

To perform a CEUI system, the emission scheme must guarantee both the decoding and unmixing of the emitted waveform portions in the backscattered signal. 
These two properties respectively ensure to obtain a satisfying spatial resolution of the output frames and to prevent from a spatial ambiguity at reception between echoes from different areas backscattered at different times of acquisition.
To address this, a specific waveform based on a continuous codded excitation method used in radar target detection \cite{malanowski_detection_2012} \cite{savci_noise_2020} is investigated.
Subsequently, a conventional pulse compression method, as introduced in \cite{cook1960pulse} for frequency modulation, to decode and unmix, is performed coupled with an appropriate sliding window approach (SWA) implemented on the recorded echoes and the continuous emission. This combination allows for the reconstruction of a sequence of 1D lines at the desired slow time frequency $f_{img}$, constrained under the sampling frequency of used by the probe $f_s$.

\section{Method}

This section investigates a method to perform CEUI, building upon the concept of continuous insonification of the medium as previously discussed. Our approach is detailed in this paper for a SISO configuration, presuming the existence of two mono-element probes capable to insonify uninterruptedly the medium while listening for backscattered echoes.

The following subsections describe the process of generating a M-mode image, $\mathbf{M}(t_E^w,z)$, for a medium containing spatially dynamic scatterers. A M-mode image serves as a spatio-temporal representation of a single-line ultrasound image, also called A-mode, throughout the acquisition time. 
%
%
The y-axis of $\mathbf{M}(t_E^w,z)$ stems for the spatial dimension represented by $z$, the axial depth of the medium, whereas the x-axis corresponds to the temporal dimension modelled by $t_E^w$, the slow time upon which 1D lines are reconstructed.

%
%
The continuous monitoring of the medium is achieved through a sequence of five steps illustrated on Figure \ref{fig:continuous_pipeline}: 
\begin{figure*}
    \centering
    \includegraphics[width=0.98\textwidth]{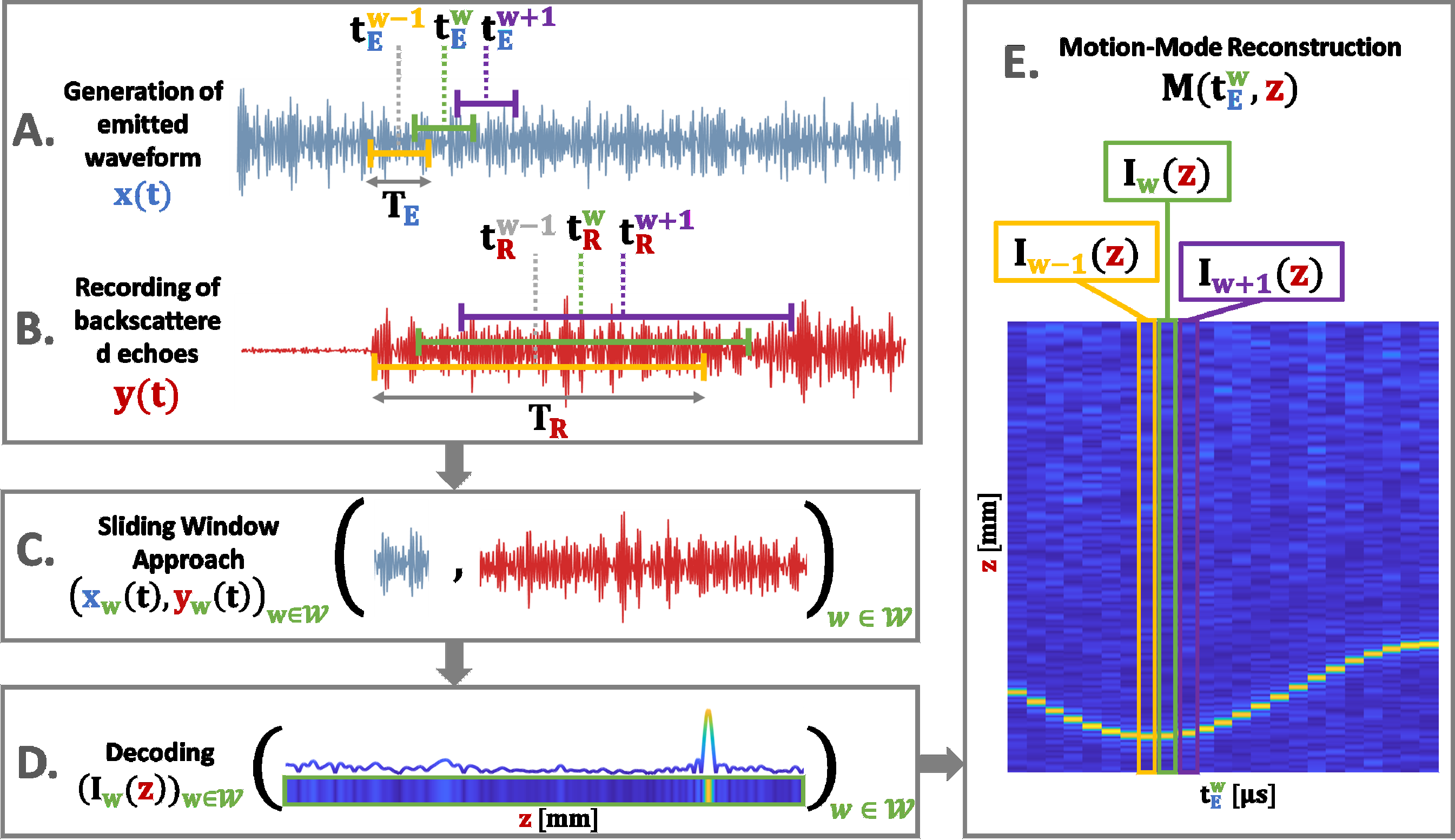}
    \captionsetup{justification=centering}
    \caption{General framework of the SISO CEUI system based on a sliding window approach. 
    Step A stands for the generation of the continuous emitted waveform $x(t)$ which insonifies the medium. %
    Then, step B refers to the reception of continuously generated echoes stored in $y(t)$.
    On step C is performed a SWA to reconstruct a set of pair of signals $(x_w(t),y_w(t))_{w \in \mathcal{W}}$ where $(x_w(t))_{w \in \mathcal{W}}$ are the reference signals and $(y_w(t))_{w \in \mathcal{W}}$ their associated backscattered echoes.
    In step D, an unmixing and decoding process using each pair introduced previously enables to build a set of 1D lines through the depth axis $\big (I_w(z) \big )_{w \in \mathcal{W}}$. Here, $z$ is function of the TOF as described latter in Eq. \eqref{Eq_scatterer_depth} in subsection \ref{subsubsection:post_processing}.
    Finally, step E illustrates the motion-mode image reconstruction $M(t_E^W,z)$ by concatenation all post-processed 1D lines along columns.}   
    \label{fig:continuous_pipeline}
\end{figure*}
\begin{enumerate}[label=\Alph*]
    \item the generation of a continuous waveform for emission (section \ref{subsec_emissionScheme}) ; 
    \item the recording of the uninterrupted backscattered echoes (section \ref{subsec_signalModel}) ;
    \item a sliding window approach on the emitted and received signals (section \ref{subsec_emission_reception_scheme}) ;
    \item a decoding and unmixing step at different instants of the acquisition to reconstruct a set of 1D lines (subsections \ref{subsubsection:MF} and \ref{subsubsection:MISMF}) ;
    \item 1D lines are post processed, then the M-mode image $M(t_E^k,z)$ of the moving medium is reconstructed (subsection \ref{subsubsection:post_processing}).
\end{enumerate}
%
%
%

\subsection{Generation of a nearly continuous imaging system}\label{subsec_emission_reception_scheme}



Consider $x(t)$ as the continuously emitted signal, and $y(t)$ as the signal continuously recorded, backscattered by the medium. From these two signals, the objective of the proposed framework is to achieve an almost continuous imaging system capable to reconstruct an image at any time with respect to the sampling frequency, noted $f_s$, of signals by the probe.

To achieve this, the current subsection exposes the overarching strategy of CEUI and provides an in-depth explanation of step C, a  key component. It aims at extracting the time windows corresponding to a specified time denoted as $t_E^w$. In the emission stage, the extraction of the time window aligned with $t_E^w$ is a straightforward process, as outlined by Eq. \eqref{Eq_signalRef}: 
%
\begin{equation}\label{Eq_signalRef}
    x_w(t) = x_{PE}(t) \cdot rect \Big ( \frac{t-t_E^w}{T_E} \Big ), \mbox{with: } x_{PE}(t) = x(t) \ast i(t),
\end{equation}
\noindent
where $t_E^w$ designates the center of the emission time window, $x_{PE}(t)$ the PE emission, and $T_E$ the length of the reference emitted signal. It is noteworthy that the tuning of $T_E$ relies on a trade-off between the shortness of the temporal scale of physical phenomenons highlighted and the accuracy of the estimated output M-mode image. 
It is necessary to convolve $x(t)$, the true waveform travelling through the medium, with $i(t)$, the piezo-electric impulse response of probe elements, like it is processed on reflected echoes.
In \eqref{Eq_signalRef}, $rect$ denotes the conventional rectangular function given by:

\begin{equation}\label{Eq_rectWin}
    rect(t) =  \left\{
    \begin{array}{ll}
        1, \; \mbox{if} \; |t| \leq 0.5, \\
        0, \; \mbox{otherwise.}
    \end{array}
    \right.
\end{equation}

At the reception stage, the time window extracted from the received signal $y(t)$ is contingent upon both the emission time $t_E^w$ and the maximal medium depth to be imaged, denoted by $R_{max}$ hereafter.
This reception time window coherent with the considered time emission, denoted by $y_w(t)$ and centered at time $t_R^w$, is given by: 
\begin{equation}\label{Eq_signalEchoes}
    y_w(t) = y(t) \cdot rect \Big ( \frac{t-t_R^w}{T_R} \Big ), 
\end{equation}
\noindent with:
\begin{equation*}
    \begin{array}{ll}
        t_R^w = t_E^w-\frac{T_E}{2}+\frac{T_R}{2} \\
        T_R = T_E + 2 \frac{R_{max}}{c},
    \end{array}
\end{equation*}
and $c$ denoting the speed of sound in the imaged medium, supposed constant and known.
By coherently extracting the emission and reception windows in this manner, it guarantees that, for any arbitrarily selected emission time $t_E^w$, the potential echoes reflected by the medium are contained by $y_w(t)$ and encoded by $x_w(t)$.
To elaborate further, the starting time of $y_w(t)$ is aligned with the one of $x_w(t)$, i.e., the first echoes may be localized at the contact with the probe, whereas the ending time depends on the starting time and the round-trip duration of the last emitted sample of $x_w(t)$ while reaching the desired maximal imaging range $R_{max}$.
The choice of the emission signal (step A) and the decoding at reception to recover the medium from the received signal (step C), are respectively detailed in subsections \ref{subsec_emissionScheme} and respectively \ref{subsec_decoding}.


As explained in the introductory section, the objective is to obtain a quasi-continuous representation of the imaged medium, $\mathbf{M}(t_E^w,r)$. A sequence of $N_{img}$ 1D lines in the depth $z$ dimension, denoted as $\big ( I_w(z) \big )_{w \in \mathcal{W}}$ with $\mathcal{W}=\llbracket 1 , N_{img} \rrbracket$ will be reconstructed. To do so, a SWA referenced as step C in Figure \ref{fig:continuous_pipeline} and based on \eqref{Eq_SWA}, is performed to extract the set of pairs of signals $\big ( x_w(t) , y_w(t) \big )_{w \in \mathcal{W}}$.
\begin{equation}\label{Eq_SWA}
    \forall w \in \mathcal{W}\backslash \{N_{img}\}, \; \left\{
    \begin{array}{ll}
        t_E^{w+1} = t_E^w + \frac{1}{f_{img}} \\
        t_R^{w+1} = t_R^w + \frac{1}{f_{img}}. 
    \end{array} \right.
\end{equation}
With the proposed framework, the slow time frequency $f_{img}$ which sets the duration separating two successive 1D lines is theoretically not bounded. However, in practice, it is constrained by the sampling frequency $f_s$ of recorded signal.
Each pair obtained using the aforementioned method is processed to decode and unmix the recorded backscattered echoes (step C) and reconstruct the M-mode image $\mathbf{M}(t_E^w,z)$ (step E) by column-wise concatenation of each estimated 1D line $I_w(z)$ (step D).

%

%
\subsection{Received signal model}\label{subsec_signalModel}

The modelisation of the interaction between a given medium and a transmitted signal is a well-established field in ultrasound imaging-related literature. However, to the best of our knowledge, all  existing works are based on the pulse echo scheme, resulting into multi-static models. This is the case, for example, for the state-of-the-art simulators in ultrasound imaging, such as Field-II \cite{jensen_calculation_1992}, K-Wave \cite{treeby_k_wave_2010}, or MUST \cite{garcia_make_2021}. In particular, existing models assume that the medium is not changing while the emitted waveform is propagating through the field of view, thus resulting into convolution models between the medium and spatially invariant or variant point spread functions. 

The proposed CEUI framework aims at going beyond this consideration, by taking account of a continuous emission interacting with a moving medium.
%
Here is assumed a bi-element ultrasound probe with an emitter at position $\mathbf{p}_{E}$ and a receiver at position $\mathbf{p}_{R}$ with no specific geometry.
We consider an imaged medium formed by $N_S$ individual point scatterers, indexed by $k$, of respective echogenicity $A_k(t)$ and position $\mathbf{p}_S^k(t)$. 
The latter parameter, in addition to the emitter and receiver element positions, enables to compute the most important feature of our model: the non stationary time-of-flight (TOF) $T_{TOF}^k(t)$ of each scatterer through the acquisition time. Indeed, it is crucial to note that scatterers motion is modeled during the propagation of the emitted wave through the medium.
%

%
\begin{figure}[H]
    \centering
    \includegraphics[width=0.45\textwidth]{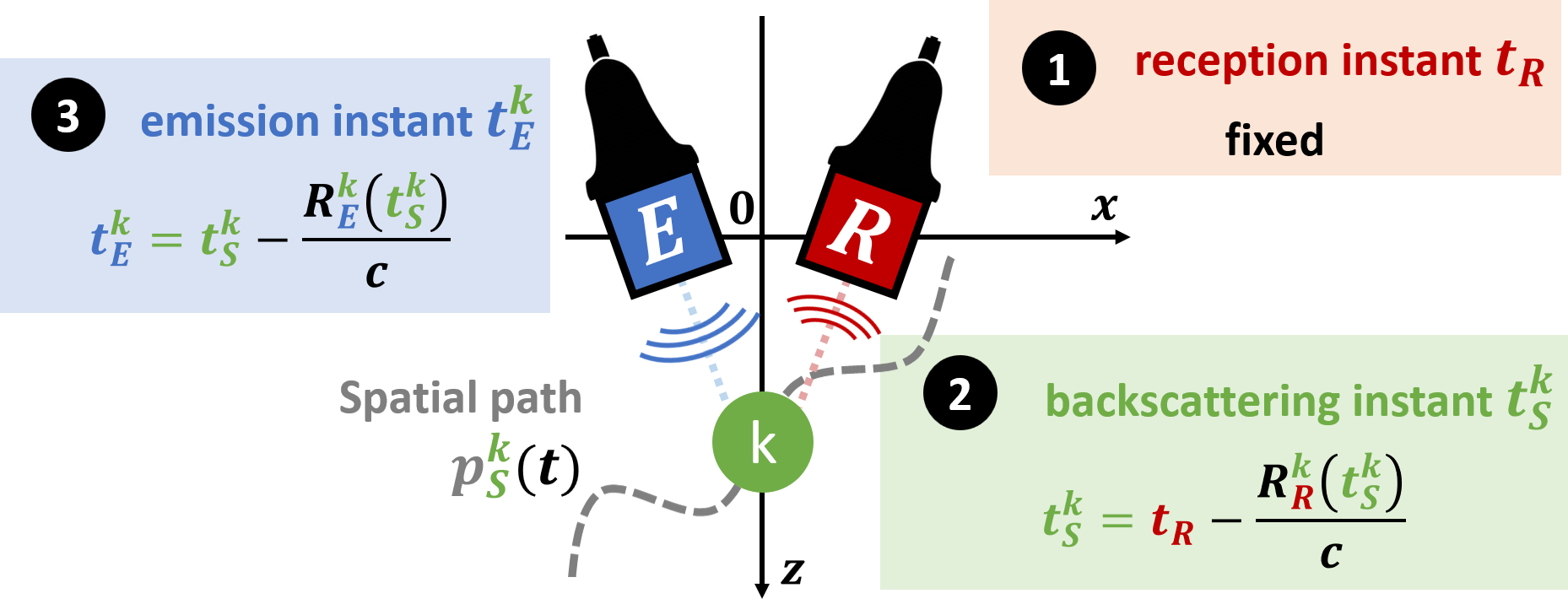}
    \captionsetup{justification=centering}
    \caption{Model used to estimate the round-trip duration of a point of an emitted waveform with a bi-element ultrasound probe, composed of a single emitting element and a single receiving element for one scatterer of indexes $k$ that composes the medium. For an arbitrary chosen reception time, $t_R$, the associated backscattering instant $t_S^k$ and the emitting instant $t_E^k$ are both estimated as suggested.}
    \label{fig:model_scheme}
\end{figure}
The proposed model and its implementation are illustrated in Figure \ref{fig:model_scheme}, for an arbitrary chosen reception time denoted by $t_R$. 
The time-of-flight $T_{TOF}^k(t_E)$ represents the time the ultrasound wave emitted at time $t_E$ needs to travel from the emitter to the reflector $k$ at its specific location at the time of interaction, and back to receiving element. It is defined by: 
\begin{equation}\label{Eq:TOF}
    T_{TOF}^k(t_R)=T_{E \rightarrow k}(t_R) + T_{k \rightarrow R}(t_R),
\end{equation}
with $T_{E \rightarrow k}(t_E^k)$ and $T_{k \rightarrow R}(t_E^k)$ respectively the round-trip times between the ultrasound probe elements and the $k^{th}$ scatterer. The latter are computed as follows:
%
%
\begin{equation}\label{Eq:TOF_details} 
    \left\{
    \begin{array}{ll}
        T_{E \rightarrow k} = t_S^k-t_E^k = 
        \frac{|| \mathbf{p}_S^k(t_S^k) - \mathbf{p}_E ||}{c} = \frac{R_E^k(t_S^k)}{c} \\
        T_{k \rightarrow R} = t_R-t_S^k = 
        \frac{|| \mathbf{p}_R - \mathbf{p}_S^k(t_S^k) ||}{c} = \frac{R_R^k(t_S^k)}{c},
    \end{array}
    \right.
\end{equation}
\noindent
where $R_E^k(t_S^k)$ and $R_R^k(t_S^k)$ are the range of the $k^{th}$ scatterer to, respectively the emitter E and the receiver R at $t_S^k$, the specific instant of interaction between the wave transmitted at instant $t_E^k$ and the k-th scatterer. These variables are used to model the relationship between the triplet $(t_E^k,t_S^k,t_R)$ on Figure \ref{fig:model_scheme}.
%
%
%
\noindent
Note that conventionally, $\mathbf{p}_S^k(t)$, the position of the scatterer, is assumed static in current models: no spatial displacement or echogenicity variation is modelled while the emitted waveform interacts with the scatterers of the medium.

Finally, by insonifying continuously a medium composed of $N_S$ point scatterers with an emitted waveform $x(t)$, the backscattered echoes regrouped in signal $y(t)$, are given by \eqref{Eq_ModeleSISO}:
\begin{equation}\label{Eq_ModeleSISO}
    y(t_R) = \sum_{k \in \mathcal{K}(t_{R})} 
    A_k(t_S^k) \cdot x (t_E^k),
\end{equation}
\noindent with $\mathcal{K}(t_R)$ the set of indexes $k$ of all scatterers insonified at their associated time $t_S^k$ leading to a reception at $t_R$: 
\begin{equation}\label{EqConditionScat}
    \begin{split}
         \mathcal{K}(t_R) = \Big \{ k \in \llbracket 1, N_S \rrbracket  \small \; | \; \exists \; t_S^k \in \mathbb{R}^{+*}, \\
        t_R = t_E^k + \frac{
        R_E^k \big (t_S^k \big )
        + R_R^k \big ( t_S^k \big )}{c}
        \Big \}
    \end{split}
\end{equation}
\noindent
The set defined in \eqref{EqConditionScat} characterizes the set scatterers that contributes to the backscattered signal at time $t_R$ and describes, for the given reception instant $t_R$, which portion of the emission is reflected and the backscattering instant $t_S^k$ which sets the roundtrip duration of $x(t_E^k)$. 
The model described in \eqref{Eq_ModeleSISO} is used to model the step B in Figure \ref{fig:continuous_pipeline}. Its implementation, assuming a linear interpolation of non integer backscattering and emitting times, is resumed in the following pseudo-code.
\begin{algorithm}[H]
    \caption{\textbf{RF signal generation for a dynamic medium assuming no attenuation}}
    \begin{algorithmic} 
        \REQUIRE $(\mathbf{p}_S^k)_{k}$, $(\mathbf{A}_k)_{k}$, $\mathbf{p}_E$, $\mathbf{p}_R$, $c$, $\mathbf{e}$,
        \ENSURE $\mathbf{y}$
        \STATE $\mathbf{y}=\mathbf{0}$
        \FOR{$k \in \llbracket 1, \; N_S \rrbracket$} 
            \STATE $\mathbf{y_k}=\mathbf{0}$
            \FOR{$n_R \in \llbracket 1:N_{echo} \rrbracket$} 
                \STATE $n_S^k$ s.t. $n_R = \frac{R_{k \rightarrow R}(n_S^k)}{c \cdot T_s}+n_S^k$ 
                \STATE $n_E^k \gets n_S^k - \frac{R_{E \rightarrow k}(n_S^k)}{c \cdot T_s}$
                \STATE $\alpha \gets n_E^k - \lfloor n_E^k \rfloor$ 
                \STATE $\beta \gets n_S^k - \lfloor n_S^k \rfloor$ 
                %
                %
                %
                %
                \STATE $\mathbf{y}{[n_R]} \gets \Bigl( (1-\beta) \cdot \mathbf{A}_k {\bigl[ \lfloor n_{S}^{k} \rfloor \bigr]} + \beta \cdot \mathbf{A}_k {\bigl[ \lfloor n_{S}^{k} \rfloor +1 \bigr]} \Bigr) \cdot
                \Bigl( (1-\alpha) \cdot \mathbf{x} {\bigl[ \lfloor n_{E}^{k} \rfloor \bigr]} + \alpha \cdot \mathbf{x} {\bigl[ \lfloor n_{E}^{k} \rfloor + 1 \bigr]} \Bigr)$ 
            \ENDFOR
            \STATE $\mathbf{y} \gets \mathbf{y} + \mathbf{y}_k$ 
        \ENDFOR
    \end{algorithmic}
\end{algorithm}

%

%
\subsection{Emitted excitation}\label{subsec_emissionScheme}

As explained previously, the proposed framework uses a continuous excitation signal $e(t)$ to insonify the medium. 
Note that the emitted signal $x(t)$ introduced in the previous section is linked to $e(t)$ through a convolution with $i(t)$, the impulse response of the piezoelectric element, introduced in Eq. \eqref{Eq_signalRef}. 
To choose an appropriate waveform, it is essential to bear in mind that the goal is to accurately reconstruct a representation of the medium at any given moment during the acquisition process.
To that end, the emission must ensure good performance in the decoding/unmixing step: the echoes from each part of $x(t)$ need to be continuously identifiable in the received signal $y(t)$. 
A straightforward choice is to use a random excitation to obtain a continuous signal with decorrelated portions and will be used in this work. More precisely, inspired by continuous wave radar applications for multi-target detection and velocity estimates, the excitation signal used herein is:
%
%
\begin{equation}\label{Eq_noiseRadar}
    \forall t \in \mathbb{R}^+, \; e(t)=A(t) \cdot cos(2\pi f_ct + \Theta(t)) 
\end{equation}
\noindent
where $A(t)$ follows a Rayleigh distribution with arbitrary parameter $\sigma$ whose value depends on the maximum energy emitted by the ultrasound element, and $\Theta(t)$ is an uniformly random distributed phase in the interval $[0,2\pi]$. $f_c$ defines the central frequency of the ultrasound probe. Based on the properties of the 2D Gaussian distribution function, in particular its rotation invariance and the independence between the magnitude and the angle of a 2D random Gaussian vector, and keeping in mind that a Rayleigh variable is defined as the magnitude of a 2D Gaussian vector, it can be shown that $e(t) \sim \mathcal{N}(0,1)$ up to a constant multiplier.


%
\subsection{Decoding}\label{subsec_decoding}

The previous subsections showed (i) the general continuous emission-reception scheme, that allows the extraction, from continuous emitted and received signals, of $x_w(t)$ and $y_w(t)$ corresponding to the medium state at a given time $t$, (ii) the excitation waveform used, and (iii) the proposed model of the continuously received signal related to the imaged medium. This subsection explains the decoding step, \textit{i.e.}, how $y_w(t)$ is combined with $x_w(t)$ to mitigate the effect of the emitted signal. More precisely, matched and mismatched filters are investigated \cite{misaridis2005use}, as described hereafter. 
In contrast to PE, where the energy of the recorded signal is mostly concentrated at times directly associated to scatterer positions, CEUI produces echoes throughout the whole acquisition time. Therefore, a decoding and unmixing process is needed to reconstruct spatial structures.

Let us consider in what follows, the sampled versions of $x_w(t)$ and $y_w(t)$, respectively denoted by $\mathbf{x}_w= \big [ x_w(T_s) \; ... \; x_w(N_E \cdot T_s) \big ] ^T$ and $\mathbf{y}_w= \big [y_w(T_s) \; ... \; y_w(N_R \cdot T_s) \big ]^T$, of respective lengths $N_E$ and $N_R$ samples. $T_s=\frac{1}{f_s}$ defines the sampling period of the imaging system.
%
%
In the following, we assume that the sampled received signal, $\mathbf{y}_w$, can be expressed as the sum between a signal of interest $\mathbf{s}_w \in \mathbb{R}^{N_R}$, representing a discrete version of the medium, and an interference signal denoted by $\mathbf{v}_w \in \mathbb{R}^{N_R}$:
\begin{equation}\label{Eq_separation_signal}
    \mathbf{y}_w = \mathbf{s}_w + \mathbf{v}_w.
\end{equation}
%
%
The discretized signal $\mathbf{s}_w$ is equal to the noise free backscattered echoes exclusively generated by the interaction between the tissues and the signal of reference $\mathbf{x}_w$. On the other hand, $\mathbf{v}_w$ contains all sources of stochastic noise, plus the echoes generated by the emitted waveforms portions of $\mathbf{x}$ apart from $\mathbf{x}_w$. These backscattering are considered as interference because they harm the estimation of the medium which interacts exclusively with $\mathbf{x}_w$.
These interference, that may harm the resulting image contrast, are mitigated by the significant higher energy contained by $\mathbf{x}_w$ compared to the one carried by a single emitted pulse in the PE paradigm. Assuming both aforementioned signals have approximately the same power, the energy ratio equals to the ratio between their temporal length.
Following the model described in subsection \ref{subsec_emissionScheme}, $\mathbf{s}_w$ can be expressed, for a SISO system, as a function of the scatterers forming the medium:
%
\begin{equation}\label{Eq_signalOfInterest}
        \begin{split}
            \mathbf{s}_w = \sum_{k=1}^{N_S} \Big \{ &
            A_k \cdot \delta[n-n_{R,first}^k] \ast
            \varphi (\mathbf{x}_w, \mathbf{n}_{\mathbf{E}}^k - \big ( n_E^w - \Big \lfloor \frac{N_E}{2} \Big \rfloor \big ) +1)  \Big \}
        \end{split}
\end{equation}
where $\mathbf{n}_\mathbf{S}^k$ are the discretized backscattering times $t_S^k$ of the $k$-th scatterer and $\mathbf{n}_{\mathbf{E}}^k$ the paired discretized emitting times $t_E^k$ for all evaluated $n_R \in \llbracket 1,N_R \rrbracket$, the discretized reception times $t_R$ and $A_k$ the echogenecity of the $k^{th}$ scatterer assumed constant.
An interpolation operator, denoted by $\varphi$, is applied on $\mathbf{x}_w$, because the latter signal is primarily evaluated at the times $\llbracket 1 \; , \; N_E \rrbracket$ but the roundtrip path and backscattering processes distort it and implies to interpolate new values at instants $\mathbf{n}_{\mathbf{E}}^k$.
The sampled instant $n_{R,first}^k$ describes the first reception instant where an echo is recorded, while $\big ( n_E^w - \Big \lfloor \frac{N_E}{2} \Big \rfloor \big )$ is the sample of the first element of the window $w$.
    
The aim of the decoding part, illustrated in the step C of Figure \ref{fig:continuous_pipeline}, is to estimate $\mathbf{s}_w$ by suppressing the information embedded by $\mathbf{v}_w$. In this work, two methods are evaluated: the conventional matched filter (MF) \cite{misaridis_use_2005} and a mismatched filter (misMF) based on the reduction of the integrated side lobe ratio (ISLR) of the point spread function (PSF) \cite{rabaste_mismatched_2015}. These filters, noted $(\mathbf{h}_w)_{w \in \mathcal{W}}$ are defined such that:
\begin{equation}
    \forall w \in \mathcal{W}, \; \mathbf{I}_w = \mathbf{y}_w \star \mathbf{h}_w
\end{equation}
\noindent where $\mathbf{I}_w$ is the discretization of a single 1D frame previously noted $I_w(z)$ and $\star$ is the cross-correlation operator.  

\subsubsection{Matched filter}\label{subsubsection:MF}

The matched filter, which simultaneously performs the unmixing and decoding of the recorded mixture of distorted echoes, is the most popular pulse compression receiving scheme. 
This adaptive filter aims at maximizing the SNR assuming additive stochastic noise and is defined by:



\begin{equation}\label{Eq_MatchedFilter}
    \forall w \in \mathcal{W}, \; 
    \mathbf{h_w} = 
    \frac{\mathbf{R}_{\mathbf{v}}^{-1} \; \mathbf{x}_w}
    { \sqrt{\mathbf{x}_w^H \; \mathbf{R}_{\mathbf{v}}^{-1} \; \mathbf{x}_w}},
\end{equation}

\noindent
where $\mathbf{R}_{\mathbf{v}}=\mathbb{E}\{\mathbf{v}_w \mathbf{v}_w^H\}$ is the covariance matrix of the theoretical noise added on the backscattering of $\mathbf{x}_w$. Assuming a white Gaussian noise, the covariance matrix is equal to the identity matrix (considering unitary variance), and the match filter becomes 
\begin{equation}\label{Eq_MatchedFilterSimplifie}
    \forall w \in \mathcal{W}, \;
    \mathbf{h}_w = \frac{ \mathbf{x}_w}{ \sqrt{\mathbf{x}_w^H \mathbf{x}_w} }.
\end{equation}

\noindent


\subsubsection{ISLR mismatched filter}\label{subsubsection:MISMF}

As introduced previously, the conventional MF assumes the knowledge of the exact pattern of the reference signal $\mathbf{x}_w$ in received echoes $\mathbf{y}_w$. However, because of the temporal non-stationarity of the medium, backscattered echoes are not only a temporally delayed version of the emitted waveform. Therefore, the decoding and unmixing method must be flexible enough to identify reasonably distorted versions of $x(t)$ as a signal of interest. Moreover, because medical imaging has to deal with complex media containing multiple echogenic structures that may be spatially close, the PSF of the imaging system must have, meanwhile, low sidelobes and a narrow mainlobe. The latter properties ensure respectively enough contrast on the image and a good separability of close objects. 

For each reconstruction of a frame $\mathbf{I}_w$, based on a variational approach developed in \cite{rabaste_mismatched_2015}, an adapted mismatched filter is designed by minimizing the ISLR of the PSF, defined by   
\begin{equation}\label{Eq_ILSR}
    \mbox{ISLR} \; (\mathbf{c}_w) = \frac{\mathbf{c}_w^H \; \mathbf{F} \; \mathbf{c}_w}{\mathbf{c}_w^H \; \overline{\mathbf{F}} \; \mathbf{c}_w} \; , \mbox{ with } \mathbf{c}_w= \mathbf{\Lambda}_K \mathbf{x}_w,
\end{equation}
\noindent where $\mathbf{c}_w$ is the PSF associated to the decoding filter and $\mathbf{F}$ is a diagonal square matrix with 1 everywhere except a 0 on the central peak of $\mathbf{c}_w$. All elements in $\overline{\mathbf{F}}$ are zero except a 1 at the position of the central peak of $\mathbf{c}_w$.
Moreover, $\mathbf{\Lambda}_K$ is a Toeplitz matrix of reverse delayed $\mathbf{h}_w$ designed to process the cross-correlation between $\mathbf{x}_w$ and the miss-matched filter $\mathbf{h}_w$ to obtain $\mathbf{c}_w$.

Using the Lagrangian multiplier approach and considering a constraint on the output energy of the filter $\mathbf{h}_w$ of length $K$, i.e., $\mathbf{h}_w^H \mathbf{h}_w = \mathbf{x}_w^H \mathbf{x}_w $, one may obtain an analytical solution to the minimization of the ISLR, given by:
\begin{equation}\label{Eq_mMF_ILSR}
    \mathbf{h}_w^{*} = \frac{\mathbf{x}_w^H \mathbf{x}_w}{\mathbf{x}_w^H \big ( \mathbf{\Lambda} _K^H \mathbf{F} \mathbf{\Lambda} _K \big )^{-1} \mathbf{x}_w}  \big ( \mathbf{\Lambda} _K^H \mathbf{F} \mathbf{\Lambda} _K \big )^{-1}  \mathbf{x}_w
\end{equation}

\subsubsection{1D-lines post-processing}\label{subsubsection:post_processing}

After decoding, an envelop detection using the Hilbert transform is performed on the absolute value of each 1D line $\mathbf{I}_w(z)$. The latter is then,
upsampled to obtain a higher axial resolution and finally column-wise stored in a matrix to reconstruct the M-mode $\mathbf{M}(t_E^w,z)$. The obtained M-mode is non-linearly spatially sampled because the scatterer depth, $z_k$, is given by:
\begin{equation}\label{Eq_scatterer_depth}
    z_k(t_s^k)=\frac{\sqrt{(T_{TOF}^k(t_R) \cdot c)^2 - \Delta x^2}}{2},
\end{equation}
\noindent
where $T_{TOF}^k(t_R)$ corresponds to the round-trip duration of the echo defined in \eqref{Eq:TOF}, $t_s^k$ the backscattering instant set by the model illustrated on Figure \ref{fig:model_scheme}, $c$ the wave celerity, and $\Delta x$ the lateral distance between the emitter E and the receiver R. A spatial resampling is therefore performed to a linear one dimensional grid.

\section{Results}\label{sec:results}
This section studies four simulation cases. The first two assess the advantages of CEUI in comparison to PE to image quick (see Subsection \ref{subsec_quick}) and fast (see Subsection \ref{subsec_fast}) events, as discussed hereafter. Then, CEUI is evaluated within more realistic simulations, with regards to the sparsity of the medium (see Subsection \ref{subsec_cyst}). Finally, a study on the potential benefits of CEUI to characterize slower phenomenon is investigated, especially with regards to the investigated depth of the imaging system (see Subsection \ref{subsec_att}). %
Additionally, Subsection \ref{subsec_simulation_parameters} provides an in-depth examination of the simulation configuration and hyperparameter tuning within the CEUI framework.
\subsection{Description of the data and simulation parameters}\label{subsec_simulation_parameters}
%
Each simulation is performed using the following configuration: two mono-element probes are modeled with an orientation of $\frac{\pi}{3}$ for the emitter and $-\frac{\pi}{3}$ for the receiver, ensuring precise imaging of the central vertical line $(0z)$ around $30$ mm. The scatterers are positioned on this line at equidistant intervals from both probes, as follows:
\begin{equation}\label{eq_coordinates}
    \left\{
    \begin{array}{ll}
    \forall k \in \llbracket 1,N_S \rrbracket , \; p_x^k=0  \\
    \mathbf{p}_R= [ \; 15 \; ; \; 0 \; ; \; 0 \; ]^T \mbox{ mm} \\
    \mathbf{p}_E= [ \; -15 \; ; \; 0 \; ; \; 0 \; ]^T \mbox{ mm},
    \end{array}
    \right.
\end{equation}
\noindent where $\mathbf{p}_E$ and $\mathbf{p}_R$ are respectively the position of the emitter and the receiver, while $p_x^k$ is the lateral coordinate of the scatterers.
%
Thus, the reconstructed 1D frames represent the medium along the depth dimension along the central vertical line $(0z)$ as illustrated in Figure \ref{fig:model_scheme}. This assumption is grounded in the belief that the directivity of both the emitter and the receiver eliminate potential scattering in other ranges beyond the line, yet still sharing the same time-of-flight in the plane $(0xz)$.

Moreover, an additive white noise is added to the recorded echoes in $\mathbf{y}$ with a SNR set at $10$ dB. The latter is computed as the power ratio between the emitted signal $\mathbf{x}$ and the noise over the bandwidth of the ultrasound probe.

To generate the recorded RF signal $y(t)$, the model outlined in the subsection \ref{subsec_emissionScheme} is employed. The use of this model enables to evaluate the capability of the CEUI system to monitor a dynamic medium with regards to the position and the echogenicity of the scatterers.
The spatial dynamic of the medium will cause a non-stationary Doppler effect on the emitted waveform taken into account by the model in \eqref{Eq_ModeleSISO} and quantified on the following Figure \ref{fig_doppler}.
\begin{figure}[H]
    \centering
    \includegraphics[width=0.5\textwidth]{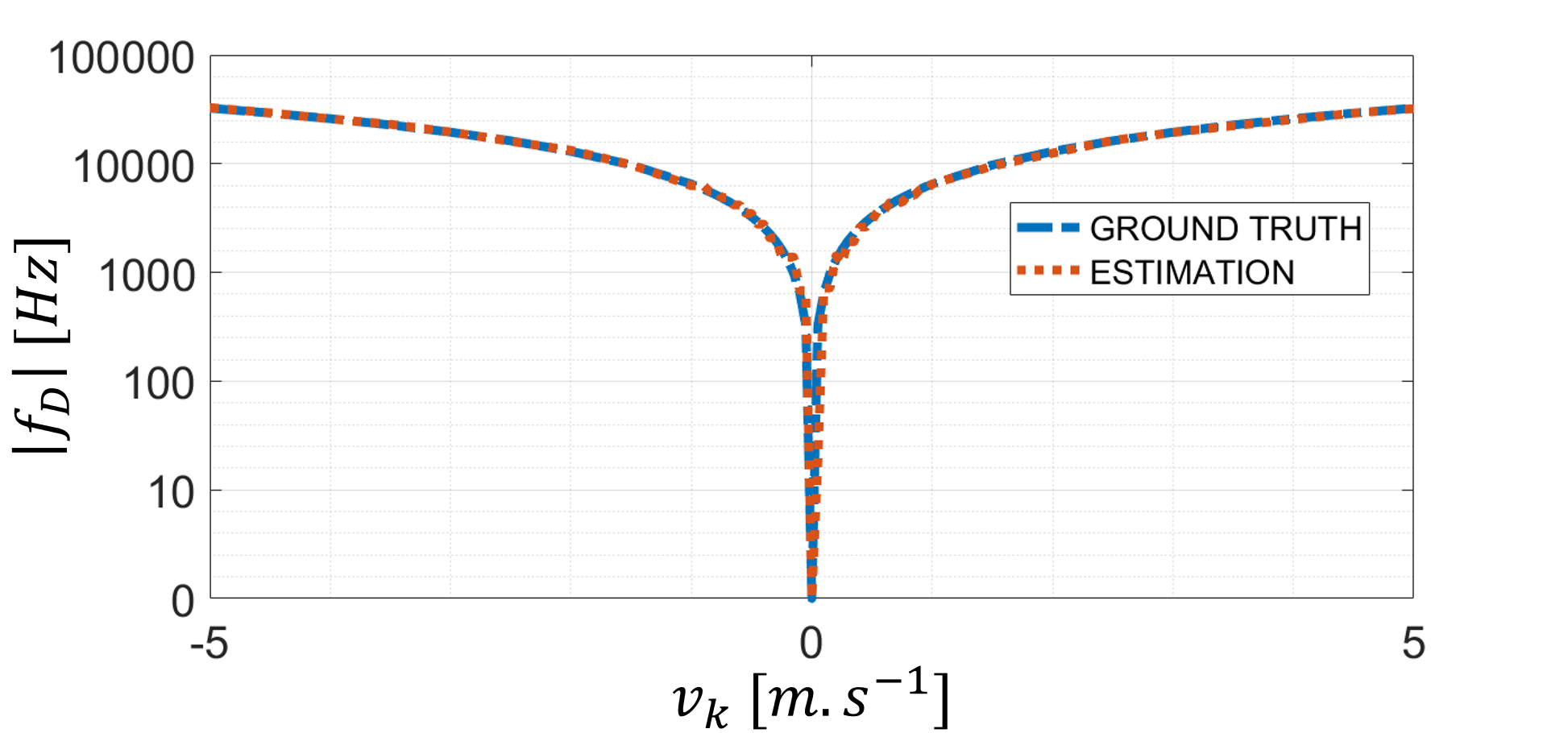}
    \captionsetup{justification=centering}
    \caption{Doppler effect engendered by a moving scatterer. A monochromatic sinus wave at $f_c=5$MHz is sent to a medium containing a single scatterer moving axially at different constant speeds $v_k$. The theoretical Doppler frequency shift in blue, is given by $f_D=f_c \cdot \frac{v_k}{c}$, with $v_k$ the scatterer radial velocity. 
    The estimated $f_D$ in orange, is obtained conducting multiple simulation, with our model, of a scatterer for each different speed $v_k$ and to quantify the frequency shift on recorded echoes.}
    \label{fig_doppler}
\end{figure}
\noindent To estimate each distortion, a simulated scatterer at a given radial velocity $v_k$ is modelled and insonified by a monochromatic emission at $f_c$. The frequency Doppler shift on the recorded echoes is then estimated for each radial velocity.
\noindent Comparing the measured distortions, in orange, to theoretical values, in blue, the spatial dynamic of scatterers is accurately modelled in our simulations.  

The simulated ultrasound probe is defined by the following parameters: a central frequency $f_c=5$MHz, a band-pass of 90$\%$ at -6dB and a sampling frequency $f_s=6f_c=30$MHz. It may be noted that the shape of the elements is not modelled. Both emitting and receiving elements, respectively located at $\mathbf{p}_E$ and $\mathbf{p}_R$, are considered as multi-directional points like transmitter and receiver. 
Indeed, for now scatterers are contained in the center of the field-of-view of both elements, therefore it is assumed that there is no distortion on the reflected echoes due to the spatial impulse response of the probe. 

The parameters of the CEUI framework are defined as follows: the reference signal, $N_E$, is set  to 251 samples ($\sim 8 \mu s$) for Subsections \ref{subsec_fast} and \ref{subsec_quick}, 2001 samples ($\sim 65 \mu s$) for Subsection \ref{subsec_cyst} and 4501 samples ($\sim 145 \mu s$) in Subsection \ref{subsec_att}. The latter parameter increases when imaged phenomenon are low and vice versa.
A step of 21 samples ($\sim 0.7 \mu s$) between successive windows is used. Therefore a slow time frequency $f_{img}$ as high as $1.8$MHz can be achieved. 
The imaged interval is set between $0$ and $4$cm which leads to $N_R \sim N_E + 1560$ samples.

The M-Mode images in Figures \ref{fig_oscillation} and \ref{fig_apparition_disparition} are shown on a narrow depth range to identify more accurately the difference of performance between PE and CEUI. 
Nevertheless, note that CEUI slow time frequency is not constrained by the imaging range, compared to PE one. 
The raw emitting signal is set as presented in the subsection \ref{subsec_emissionScheme}.

As a reference, M-mode images obtained using a CEUI approach will be compared to M-mode images reconstructed using a PE approach. To do so, a set of pulse emission of identical 13-bits Barker codded emission \cite{zhao2007barker} insonifies the medium. The latter are temporally spaced of $\frac{2 \cdot R_{max}}{c}$, the pulse repetition interval, providing a framerate of $f_{img,PE}=19kHz$ considering too, an imaging range from 0 to $4$cm and an homogeneous wave velocity throughout the medium $c=1540 \; m.s^{-1}$. This is a 100 times smaller compared to CEUI slow time frequency.
The associated decoding method for PE is matched filter.

\subsection{Fast dynamic event imaging}\label{subsec_fast}
%
The continuous interaction with the medium and a sufficient imaging framerate are two necessary conditions to extract a a correct spatio-temporal representation of a non-stationary medium. These factors ensure, if they are well tuned, first, to catch all the necessary information about the medium anywhere in the field of view at any time of the acquisition. Second, the latter recorded data can be processed, such that the imaging method get rid of a potential stroboscopic effect. The next simulations aim at evaluating the robustness of the CEUI approach to fulfil the aforementioned conditions. 

\begin{figure}[H]
    \centering
    \includegraphics[width=0.5\textwidth]{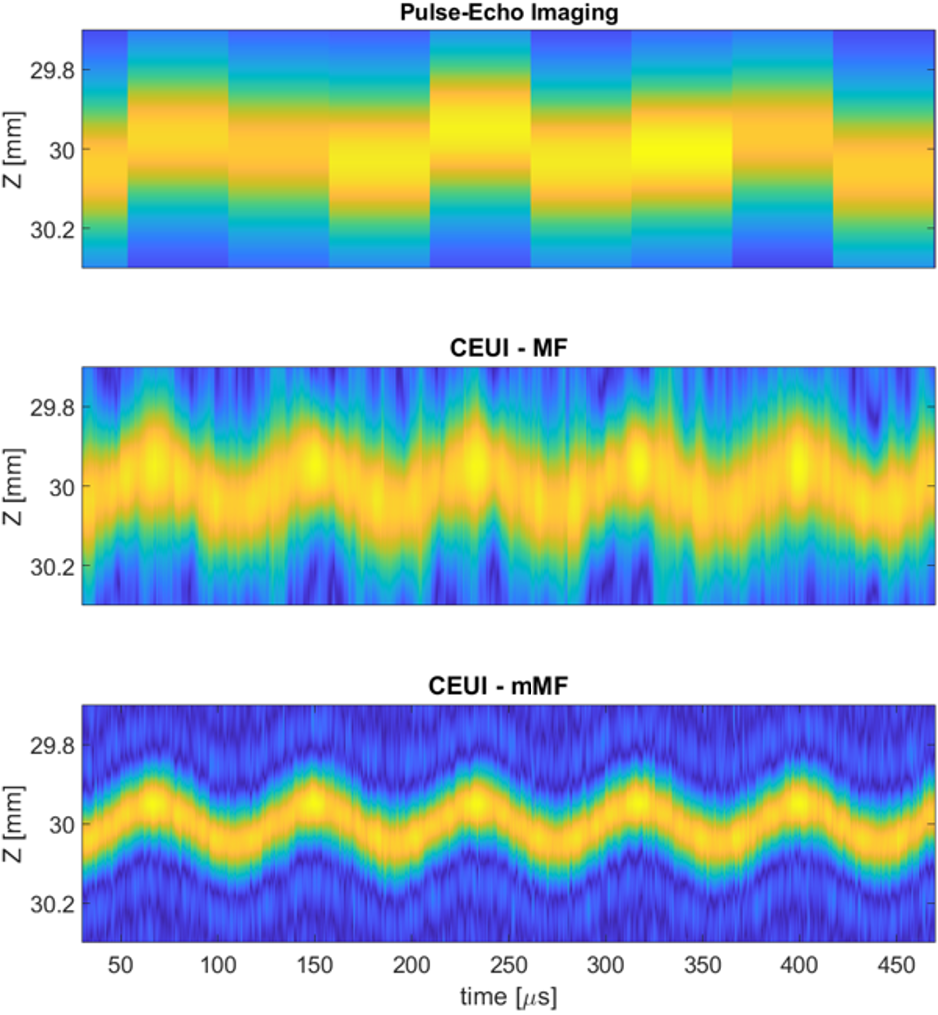}
    \captionsetup{justification=centering}
    \caption{M-Mode imaging of an axially oscillating scatterer at $12$kHz around $30$mm. The output image using PE at a slow time frequency of $19$kHz is displayed, while CEUI at a slow time frequency of $1.4$MHz is showed in the middle, respectively below with MF, respectively mMF as decoding methods.}
    \label{fig_oscillation}
\end{figure}

A first simulation is performed to challenge the true imaging framerate the CEUI can reach through an example of a vibrating echogenic scatterer imaged on Figure \ref{fig_oscillation}.
%
%
A medium containing a single scatterer axially oscillating around $30$mm depth at a frequency $f_{osc}=12$kHz $\sim \frac{2}{3} f_{img,PE}$ and peak-to-peak amplitude of vibration equals to $0.1$mm. 
Therefore, in these conditions the sinusoidal profile of the velocity varies between $[-3.8,+3.8] \; m.s^{-1}$ in the axial direction. 
%
The parameters presented previously in the subsection \ref{subsec_simulation_parameters} are used: $N_E=251$ samples is approximately equal to $\frac{f_s}{10 \cdot f_{osc}}$ to catch coherent phenomenon in all windowed recorded echoes $(\mathbf{y}_w)_{w \in \mathcal{W}}$. Indeed, if $N_E$ is not sufficiently short, each reference signal $\mathbf{x}_w$ will interact with the scatterer at too many different ranges which will lead to a poor image axial resolution. 

On the M-mode images produced using CEUI approach, all periods of the scatterer oscillation are clearly depicted because, first, the recorded RF data contain echoes from each oscillation phase within the acquisition time. Secondly, the decoding step, using both misMF and MF, remains robust to deformations of echoes due to high velocity motion. Because of the SWA, a spatio-temporal spreading of the highlighted phenomenon along both x and y-axes of the M-Mode is engendered. 
Specifically, considering $\mathbf{x}_w$ as the reference to estimate a line of the medium, it interacts with the scatterer at different positions which widens the spot vertical on top of the impact of mainlobe width of the decoding filter PSF. This impact of the $N_E$ coupled with the scatterer velocity can not be prevent when decoding, as $N_E$ cannot be as short as desired to ensure good performances.
%
In contrast, as expected for the PE approach, a stroboscopic effect is observed on the imaged motion. In the corresponding M-mode image, the five complete periods of vibration fail to be distinctly highlighted in yellow. This arises from two primary factors: first and foremost, the slow time frequency of the PE M-Mode $f_{img,PE}$ does not fulfill the Shannon theorem. Specifically, the latter is more than twice smaller than the maximal frequency of the scatterer depth position which is $f_{osc}$. Secondly, even though the true scatterer depth at the interaction instant with the emitted pulse is well estimated with PE,  RF data lack information about the medium at instants other than those during the interaction of each of the seven emitted pulses.  

%
The use of the misMF enhances drastically the axial resolution of the image compared to MF: misMF decoding provides a half-power mainlobe of $1.1 \lambda$ compared to $2.4 \lambda$ and $2.5 \lambda$ using MF decoding respectively with a CEUI and a PE approach. Here, $\lambda$ corresponds the wavelength of the central frequency $f_c$ of the probe. 
%
The contrast of the image around the true scatterer position is also improved by the misMF decoding as the mean ISLR, as depicted in \eqref{Eq_ILSR}, in the optimized ranges (all backscattering containing a portion of the signal of reference $\mathbf{x}_w$) of the M-mode is 3 times smaller than using MF for CEUI. A mainlobe width of $\frac{\lambda}{2}$ is considered to compute the ISLR. 
%
On the other hand, a small loss of performance regarding the same criterion is observed on far fields from the scatterer with a loss of 10$\%$ using misMF compared to MF in CEUI. Note that these areas are not displayed on the Figure \ref{fig_oscillation}. 
Consequently, the use of misMF for decoding increases the solubility in complex medium imaging problems with spatially close echogenic structures. However, the image contrast will not be necessarily improved as it is evidenced by an the unchanged performance with regards to the ISLR and the Peak to Sidelobe Ratio (PSLR) on the whole M-mode images. PE approach provides the main advantage that almost no decoding artifacts appears on the M-mode image thanks to the narrowness of the temporal support of the emitted waveform.   

\subsection{Quick appearing event imaging}\label{subsec_quick}

\begin{figure}[H]
    \centering
    \includegraphics[width=0.5\textwidth]{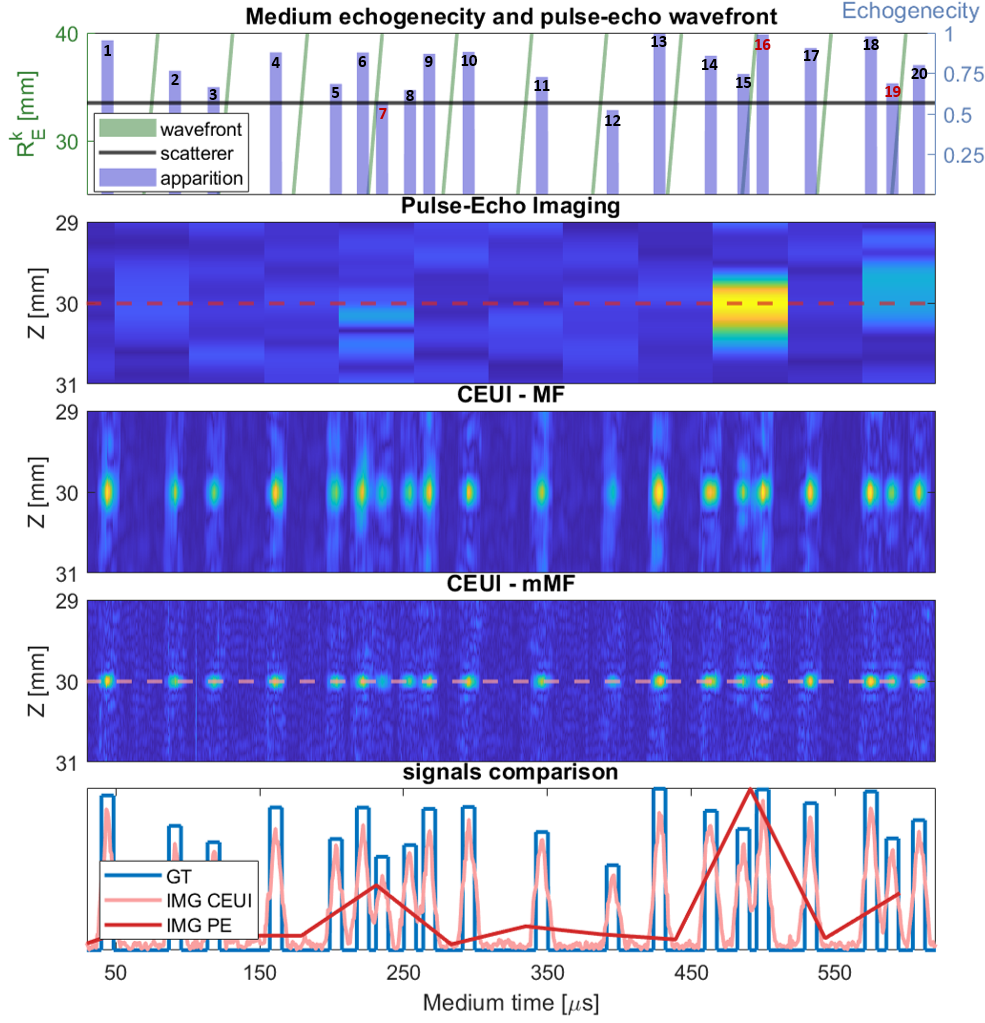}
    \captionsetup{justification=centering}
    \caption{M-Mode images of a spatially static blinking scatterer located at $30$mm depth. The blinking instants are randomly distributed during the whole acquisition. The echogenecity $A_k$ is also uniformly random distributed between 0.5 and 1.
    The first graph displays, in blue the periods when the scatterer is echogeneous, in green the ranges insonified by the emitted waveform by PE, and in gray, the horizontal line is located at the depth of the scatterer.
    The first M-mode (top to bottom) image results from a PE approach at $19$kHz, while the second and third ones use a CEUI approach at $1.4$MHz with respectively the MF and the misMF as decoding techniques.
    The fourth figure plots the ground truth M-Mode line at 30mm depth and its estimation in pink, respectively red, using CEUI respectively PE.}
    \label{fig_apparition_disparition}
\end{figure}

The simulation illustrated on Figure \ref{fig_apparition_disparition} highlights the main contribution of the continuous insonification while using the CEUI approach: because the continuous emitted waveform and the echogenic structure in the field of view are interacting without interruption, our imaging system is capable to capture short phenomenons lasting only a dozen of $\mu$s.
%
The same probe and post-processing settings as described previously are performed for a simulated medium containing a spatially static blinking scatterer at $30$mm depth with a randomly distributed echogenicity at each apparition.  

The upper graph on Figure \ref{fig_apparition_disparition} displays the ground truth blinks of the scatterer during periods in blue. The horizontal grey line represents the scatterer depth at 30mm, while the green areas are the wavefronts of the successive pulse emissions of PE approach. When a green beam meets the blue area at the scatterer depth, it means the current pulse is at least partially backscattered which permits the detection: this is the case for the $5^{th}$ and $9^{th}$ scatterer apparitions. No backscattering is produced during other apparitions of the scatterer. The difference of amplitude between the two detected blinks depends on the proportion of the emitted pulse which interacts with the scatterer.
%
The improper blinking identification using PE leads to a bad interpretation of the events occurring within the medium. Around $200 \mu$s, multiple instances of blinking (7, 16 and 19) are confused with a single occurrence, or conversely, situations where an actual appearance takes place go unnoticed (every other blinks). 
%
It is concretely shown by the red signal of Figure \ref{fig_apparition_disparition} fourth plot where only 3 echoes are recorded out of 20 potential ones. Under PE paradigm, blinks identification relies on multiple factors which are: the apparition frequency, their duration and their temporal localization.
%
However, even when apparitions are detected by PE, the proportion of pulses interacting with the scatterers can introduce a bias in the relative echogenecities imaged in the M-Mode. 
For instance, the 7th blink interacts only with the tail end of the pulse, resulting in a falsely lower representation compared to the 15th blink.  

While PE reveals limitations to detect brief events with regards to the pulse repetition interval, the two lower M-mode images on Figure \ref{fig_apparition_disparition}, obtained using CEUI, depict well all the 20 blinks. 
RF data can record echoes regardless of the duration of the scatterer’s appearance. The accurate identification of these phenomena then hinges on their duration and the intervals between them at the same depth, both intricately linked to the tuning of the parameter $N_E$.
%
%
%
%
In addition, the intensity contrast difference between blinks is generally well depicted on both CEUI M-Mode images thanks to a complete interaction during each apparition period.
%
%

%

%
%
\subsection{Cyst imaging with speckle}\label{subsec_cyst}
\begin{figure}[H]
    \centering
    \captionsetup{justification=centering}
    \includegraphics[width=0.49\textwidth]{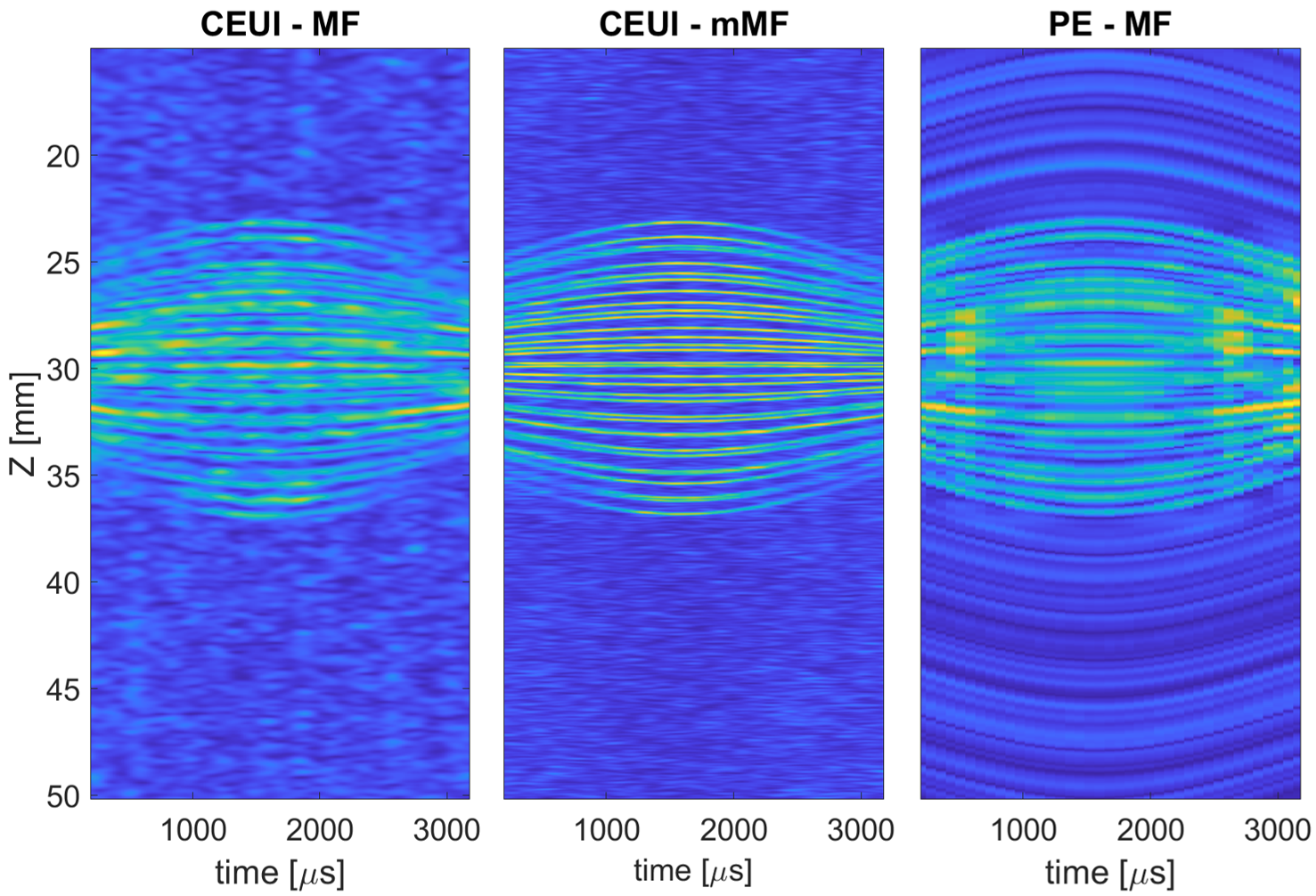}
    \caption{M-Mode images of dilated and compressed cyst surrounded by a background. The cyst is modelled by 50 scatterers axially spaced, in average, by $\lambda$ with an average echogenecity of 1. The background contains 10 scatterers randomly set every $\lambda$ in depth with an echogenecity defined by a Rayleigh distribution centered in 0.05. The total number of background scatterers reaches 950.}
    \label{fig_cyst}
\end{figure}

Due to the continuous insonification of the medium, in a more realistic scenario, the emitted waveform propagates through a medium, generating low-energy echoes that contribute to the speckle in the image.
%
%
The following experiment aims at evaluating the capacity of CEUI to identify a cyst within a material background whereas the energy of interference signal $\mathbf{v}_w$ is significantly more important than in experiments led in subsections \ref{subsec_fast} and \ref{subsec_quick}. This is due to the increased number of echogenic cyst scatterers plus the presence of a background. The simulated M-Mode images are represented in Figure \ref{fig_cyst}.

Note that $N_E$ was increased in this experiment from 251 samples to 2001 samples, to increase decoding performances in the presence of a slower moving big structure. However, it does not impact the slow time frequency of the resulting M-Mode which remains at $1.4$ MHz. Thanks to this frequency much higher than the bound imposed by Shannon theorem, it is even possible to perform temporal compounding using neighbor 1D-lines to improve image quality. For illustration purpose, a hamming window was applied in the summation of the 1D-lines to preserve temporal resolution.
Qualitatively, both decoding methods applied using CEUI enable clear identification of the central dilating cyst. The boundaries of the object are distinctly depicted. The deeper the scatterers in the cyst, the less they move which therefore, increases decoding performances of their echoes.  For both PE and CEUI, the maximal dilatation is better depicted because the medium reaches a zero velocity state. Quantitatively, the contrast between the cyst and the background increases the SNR of $3$ dB, (respectively remains the same) using mMF (respectively MF) as CEUI decoder compared to PE.
\subsection{Contrast improvement in a medium with attenuation}\label{subsec_att}
The current subsection investigates the pertinence of the use of CEUI to image a slower medium. To do so the peak signal-to-noise ratio (PSNR), as indicator of the image contrast, will be locally quantified on M-Mode images of a linearly attenuated medium defined by 6 static scatterers. The attenuation is modelled, similarly to ultrasound simulators such as Field II \cite{jensen1999linear}, by:
\begin{equation}\label{Eq_attenuation}
    A_{att}^k(R_{E\rightarrow k},R_{k \rightarrow R}) = exp \big( -\alpha \cdot f_c\cdot (R_{E\rightarrow k}+ R_{k\rightarrow R}) \big ) \\ 
\end{equation}
with $\alpha=1.5$ dB/MHz/cm. The scatterers are placed at depths of 30, 45, 60, 75, 90 and 105 mm, with the same constant echogenecity through the acquisition time. 

\begin{figure*}
    \centering
    \captionsetup{justification=centering}
    \includegraphics[width=0.95\textwidth]{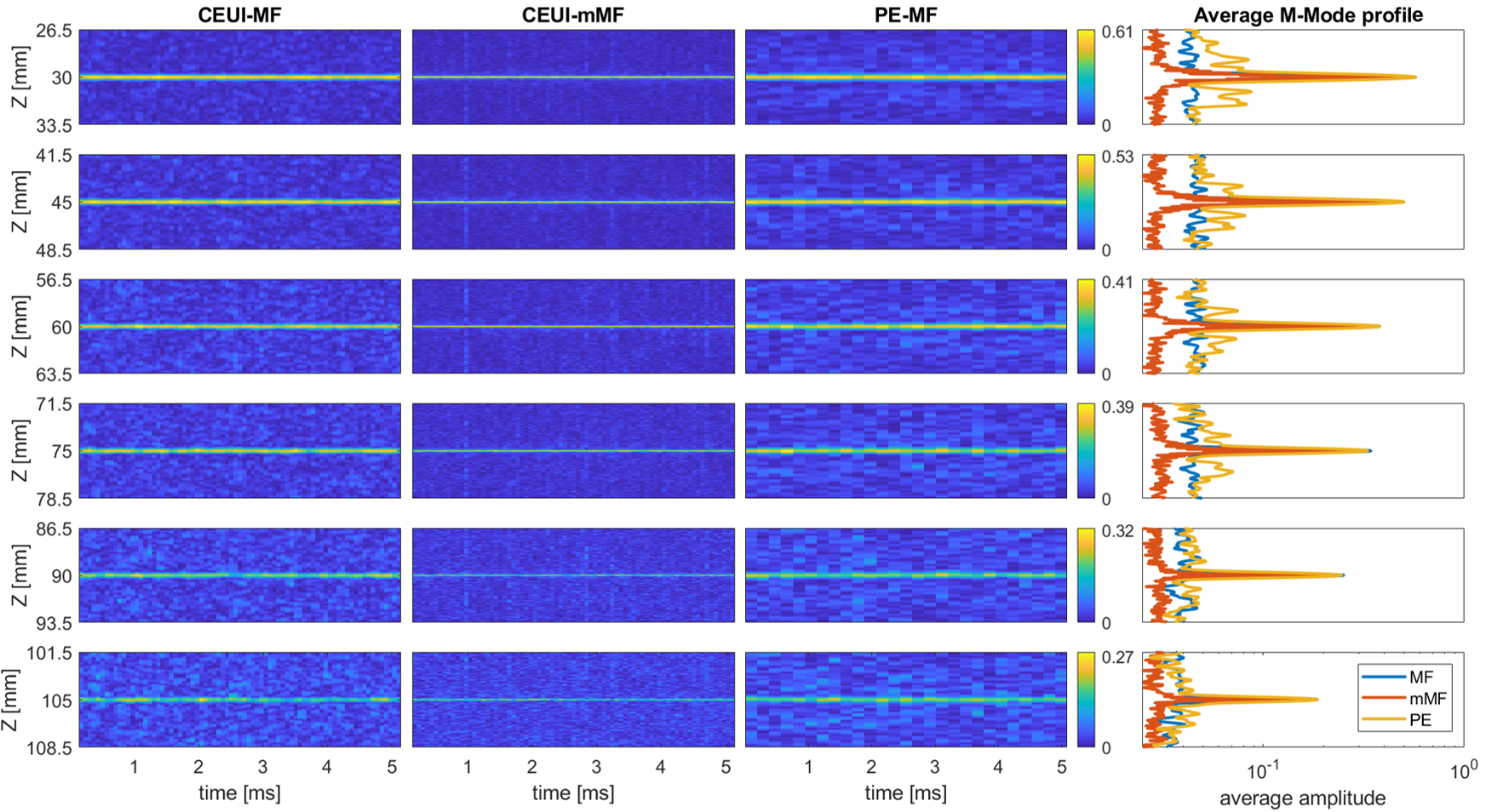}
    \caption{Regions of interest on M-Mode images built of a medium composed of 6 static scatterers with an equal static echogenecity during all the acquisition time. A linear attenuation due to the propagation of the emission through the medium is incorporated with a coefficient $\alpha=1.5$dB/cm/MHz.
    From left to right, the methods used are CEUI with MF decoding, CEUI with mMF decoding and finally PE with MF based pulse compression. The last column images show the average profile obtained using each method at each depth.}
    \label{fig_attenuation}
\end{figure*}

We assume the most challenging case, akin to subsection \ref{subsec_fast}, in which the medium can move with a maximal velocity of $v_M = 1 , \text{m/s}$ and a minimal amplitude of $\Delta z = 0.1 , \text{mm}$. From this, we can define the reference signal width to the following limit:
\begin{equation}
    N_E \leq \frac{\pi \cdot \Delta z \cdot f_S}{v_M} \sim 4501 \mbox{ samples}
\end{equation}

The resulting M-Mode images are regrouped in Figure \ref{fig_attenuation}. One can note that PE generates higher main-lobe peaks, but with significantly higher sidelobes. Computing the average over slow time of the PSNR on each 1D-line around each scatterer depth, quantifies the image contrast.
CEUI-mMF improves the PSNR by 3 to 5 dB depending on the image depth. Consequently, still based on the assumption of a linearly attenuated medium, and considering slower phenomenon to enable a wider reference signal, improves the investigated depth by 2 cm without compromising the temporal resolution of the acquisition.

%

% ...
\section{Discussion}\label{sec:discussion}
\subsection{Results interpretation}
%


Some interesting properties of CEUI are brought to light by the two proposed simulation cases in Section \ref{sec:results}.
%
%
On the one hand, the blinking scatterer case presented in Figure \ref{fig_apparition_disparition} illustrates well the capability of CEUI to catch short duration events of few $\mu$s like an echogenic object which does not remain in the imaged section of the medium: for instance, a blood vessel that is not exclusively located in the plan $(0xz)$ will contain echogenic elements that propagate through the blood flow will cross the imaging section. That generates a succession of apparitions and disappearance on a short amount of time. 
Thanks to the sufficient shortness of $\mathbf{x}_w$ and the high enough slow time frequency, the decoding step catches each single blink on the M-mode images using CEUI approach. For several successive windowing $w$, their associated reference signal $\mathbf{x}_w$ contains a portion of the backscattered emitted waveform identified in $\mathbf{y}_w$. Nevertheless, $N_E$ is small enough to limit the temporal sprawl of the apparition spots along the x-axis of the M-mode, which can prevent from temporal separability of events at same range.    

On the other hand, the oscillating scatterer case showed on Figure \ref{fig_oscillation} is relevant of several strengths of CEUI. 
CEUI generates motion-mode images that depict accurately all the oscillation periods thanks to a sufficiently high imaging framerate and a continuous interaction. 
The scatterer velocity varies between $[-3.8 \; , \; 3.8] \; m.s^{-1}$ which generates a Doppler shift of $\pm 12$kHz on the recorded echoes $\mathbf{y}$. Despite this distortion, MF and misMF based decoding methods performances are not deteriorated (mainlobe width and image contrast for instance). The robustness of the decoding method to fast moving mediums depends mostly on the tuning of $N_E$: ideally, a single portion $\mathbf{x}_w$ must interact with a version of the medium which remains approximately the same, otherwise scatterers spots will dilate along the spatial axis of the motion-mode image.  
However, if the medium of interest is assumed at a lower maximal velocity and variation, it is possible to improve PSNR and to increase the actual investigated depth using a larger value for $N_E$. With an adequate tuning to respect Shannon theorem considering a priori knowledge on involved phenomenon to be imaged, decoding performances are improved as well as the robustness to noise.

In opposition, PE using Barker codes is not capable to catch every blink or oscillation period even if the emitted waveform is longer than a conventional pulse. This inability results from the roundtrip duration based reconstruction method associated to PE and the need to prevent from an ambiguity about the origin of the echoes. However, it provides a better image contrast thanks to the lack of interference present with CEUI between $\mathbf{x}_w$ and $\mathbf{s}_w$.
As shown on Figures \ref{fig_oscillation} and \ref{fig_apparition_disparition}, the main interest of CEUI compared to PE lies in the fact that no information about the medium is missing in the recorded echoes $\mathbf{y}$ and the targeted events can be highlighted to some extent if the hyper-parameters of off-line framework are adequately tuned. 

Finally in both cases, vibrating and blinking scatterer, a better resolution in the depth dimension $z$ of the M-mode image is obtained by decoding with the misMF as its frequency content is wider than one of MF, which narrows the mainlobe.


While ultrafast ultrasound 2D imaging approaches reach an imaging framerate of an order of magnitude of $1$kHz using 15 compounded plane waves (for a $4$cm maximal imaging depth)  as presented in \cite{montaldo_coherent_2009}, CEUI has the potential to increase significantly the temporal resolution in plenty of ultrasonic applications. It is crucial to note  that the extension of CEUI from 1D line monitoring to 2D imaging will not reduce the imaging framerate it can reach because our method does not depend of the number of piezo-electric elements used for emission and reception.
Besides, in a sense, CEUI emission is comparable to PE approaches using long codded excitation such as frequency modulated waveforms or pseudo-random binary codes. The latter increase the transmitted energy to the medium and therefore, the SNR of reconstructed images \cite{tiran_multiplane_2015}\cite{pialot_sensitivity_2022}\cite{tamraoui_complete_2023} by sending insonifying longer excitation. Similarly, CEUI emission enables also to transmit larger energy considering a longer signal of interest $\mathbf{x}_w$.

Finally, thanks to the continuous insonification of the medium, assuming necessarily that the region of interest remains in the overlapping area of the fields of view of both emitter and receiver elements, the information about all the medium is captured continuously as mixture of non temporally coherent echoes. 
The robustness of the CEUI framework has been evaluated on fast and short spatial events and proves the capability of the off-line framework to reconstruct a 1D monitoring of the medium.

\subsection{Limitations and improvements}
%
%
%

However, by continuously insonifying the medium, the generated echoes $\mathbf{s}_w$ from $\mathbf{x}_w$ are noised by $\mathbf{v}_w$, the echoes generated from other portions of the emission. However, the advantage is that, assuming small Doppler effects due to scatterers motion, the content of $\mathbf{v}_w$ is tractable so, optimized emitted waveforms and decoding strategy can be explored to reduce these interference. It would be interesting to add a prior information in the decoding step to take account of prior and posterior emitted waveform portions that may interfere with $\mathbf{s}_w$. 

Depending on the tuning of $N_E$ and $\Delta t = \frac{1}{f_{img}}$, the M-mode image of the medium highlights the dynamic phenomenons at a desired temporal scale. For instance, if $N_E$ was bigger to produce the M-mode images on Figure \ref{fig_apparition_disparition}, two close blinks events could have been confused as a single and longer blink. 
%
%
Similarly, the oscillations on the left image on Figure \ref{fig_oscillation}, produced using PE, can only highlight the true oscillation of the scatterer if the Shannon theorem is respected: $N_E \leq \frac{f_s}{2 \cdot f_{osc}}$. It must be even smaller to detail more accurately the spatial path of the oscillating object. However, using a too small reference window size shortens the temporal scale of highlighted events, but considering the current decoding and unmixing methods, the variance estimation increases and the robustness to distortions in echoes decreases.

More broadly, M-mode images presented in the result section work as a proof of concept and a display of the potential of the CEUI scheme for increasing the data acquisition rate, compared to conventionally used PE based approaches, in a simple setup. Nevertheless, the simulations carried out for now are not yet realistic to some extents. First and foremost, only sparse media have been investigated, which is not realistic with regards to the complexity of media in medical applications.

The model to generate ultrasound data rely on multiple simplifying assumptions: only the piezo-electric impulse response of the probe is modeled. It would be more realistic to incorporate the directivity of the probe elements and their spatial impulse response. The factors would incorporate more variability on the backscattered echoes and potentially complicate the image reconstruction because the decoding step relies essentially on the similarities between $\mathbf{h}_w$ and $\mathbf{y}_w$. 

However, the main improvement will be to perform CEUI using a multi-element probe to extend and adapt the current framework to reconstruct a set of 2D images. The signal model to generate simulated data in a MIMO (multi input multi output) configuration is presented in the Appendix I and will be the subject of further studies.
\subsection{Potential applications}
%

Thanks to the increased temporal resolution, CEUI has the potential to benefit to a lot of applications. For instance, it would permit to reduce the time of acquisition of ULM (Ultrasound Localisation Microscopy) where the patient must remain static for several minutes in order to acquire a sufficient number of frames to reconstruct a reliable network of blood vessels \cite{hingot_microvascular_2019}. Indeed, as many micro-bubbles will be imaged in shorter amount of time if CEUI is performed in 2D. 

CEUI can also improve echocardiography methods for blood flow velocity estimation, which separated in two main approaches, respectively Pulse Wave Doppler (PWD) and Continuous Wave Doppler (CWD). 
PWD estimates the blood flow velocity at a given localization \cite{jiang_measurement_2008} but at low temporal resolution even in using high-framerate methods \cite{poree_high-frame-rate_2016} \cite{hasegawa_high-frame-rate_2011}, while CWD enables to access to a velocity quantification \cite{brown_use_2002} at any time but without axial localization.
CEUI could be an emission scheme used in this application to extract a spatial velocity map at high temporal resolution by identifying the origin of the echoes thanks to our encoding \ref{subsec_emissionScheme} and by analyzing the Doppler spectral shift at each depth.     
%

Regarding the potential improvement CEUI approach can provide for diverse ultrasound applications, to extend the 1D imaging scheme to a 2D imaging system using a multi-array probe would be a priority to investigate. 
An implementation based on the SISO CEUI framework combined with an approach similar to a full-matrix STA, a pulse compression and a temporal compounding, may be intended.
For this purpose, the development of, a set of quasi-orthogonal emission waveforms with a beamforming method which will be capable to unmix and decode the contribution of each portion of each continuous emission on each recorded signal, will be investigated.

As mentioned in the introduction, the primary motivation of our proposition, the CEUI paradigm,  aims in the end, to accelerate 3D ultrasound imaging. Despite all efforts to accelerate the acquisition, similar performances to 2D imaging are still not achievable. Even if not demonstrated here, the simple proof of concept in 1D clearly shows a potential increase in acquisition rate of at least one order of magnitude which could definitely benefit 3D ultrasound imaging especially for cardiovascular applications. 
Note also that not only the acquisition time should be reduced, but also the sensitivity for the whole imaging setup should be increased. Combined with transducers with better sensitivity, such as Capacitive Micromachined Ultrasonic Transducer (CMUT) \cite{caronti_capacitive_2006}, we could also anticipate important penetration and resolution improvements.  
Proper evaluation of safety limits should also be performed. Indeed when transmitting in continuous mode, their risk of heating and as a consequence damaging the imaged tissue increases also \cite{bhatti_comparative_2022}.

The concept needs of course to be validated in simulations, and then specific material should be designed and produced to perform the \textit{in vitro} and then \textit{in vivo} experiments.
Finally the clinical benefit of such an imaging mode will need to be demonstrated. All these aspect will be further developed and are well beyond the scope of this first proof of concept.   
\section{Conclusion}\label{sec:conclusion}
Continuous Emission Ultrasound Imaging (CEUI) enables to monitor a single line within a rapidly moving medium in the form of a M-Mode image featuring a  significantly larger slow time frequency compared to pulse-echo based conventional US imaging techniques (100 times larger). A simulation study, realistically modelling the medium motion as the waves interacts with the echogenic regions, has been conducted. In each case, CEUI demonstrates superior reactivity and robustness to short and fast events, outperforming ultrafast pulse-echo emission using Barker encoding.

The continuous insonification of all the medium in the field-of-view at any acquisition time enables to record, without aliasing, even high velocity phenomenons. This occurs to be beneficial to echocardiography, ultrasound localisation microscopy, and ultrasound imaging more generally by suppressing the period of blindness during the acquisition. The latter point paves the way for potential novel methods exploiting the temporal coherency embedded by radio-frequency data. 
  
Future studies will be carried out to, first, extend the current framework to a highly temporally resolved two and three-dimensional monitoring of dynamic medium. This will raise some challenges both for the generation of adequate and optimized continuous waveforms and for the reconstruction step and will rely on the signal model in Appendix I.
In parallel, \textit{in vitro} and \textit{in vivo} experiments protocols will be developed to assess the feasibility of the method and to evaluate its performances on real data.
 
Like in sonar and radar applications, CEUI proposes to get rid of the pulse-echo paradigm which, intrinsically constraints echography for being capable to catch quicker phenomenons characterization and to reach better temporal resolutions. 

\appendices


\section{MIMO Signal Model}\label{Annex_MIMO}

The times $t_E^{i,j,k}$ and $t_R^j$ are respectively the emission time and the reception time. Furthermore, $t_S^{j,k}$ defines the instant when an ultrasound wave is backscattered by scatterer $k$ so that it reaches element $j$ at instant $t_R^j$.

Note that $t_S^{j,k}$ only depends on the scatterer and receiving element indices, but not on the emitting element. Indeed, the backscatterered wave may come from any emitting element. Once $t_S^{j,k}$ is obtained, $t_E^{i,j,k}$ is estimated for each $i$ emitter.

Finally, the signal received by any element $j$ at time $t_R^j$ can be modelled by:
\begin{equation}\label{Eq_ModeleComplet}
    \forall j \in \llbracket 1,N_{R} \rrbracket, \;y_j(t_R^j) = \sum_{i=1}^{N_{E}} \sum_{k \in \mathcal{K}_{j}(t_{R})} 
    \Big \{ A_k(t_S^{j,k}) \cdot
    x_i (t_E^{i,j,k}) \Big \} 
\end{equation}
%
%
%
%
\noindent with $\mathcal{I}_j(t_R^j)$ the set of indexes $k$ of all scatterers insonified at their associated time $t_S^{j,k}$: 
\begin{equation}\label{EqConditionScat_MIMO}
    \begin{split}
         \mathcal{K}_j(t_R^j) = \Big \{ k \in \llbracket 1, N_S \rrbracket  \small \; | \; \exists \; t_S^{j,k} \in \mathbb{R}^{+*}, \\
        t_R^j = t_E^{i,j,k} + \frac{
        R_E^{i,k} \big (t_S^{j,k} \big )
        + R_R^{j,k} \big ( t_S^{j,k} \big )}{c}
        \Big \}
    \end{split}
\end{equation}
Note that \eqref{EqConditionScat_MIMO} is obtained from \eqref{Eq:TOF_details}, assuming that scatterers move at subsonic speeds, which in turn results into only one triplet $\big ( t_E^{i,j,k},t_S^{j,k},t_R^j \big )$ for any $k$.
In other words, the signal received by element $j$ at time $t_R^j$, is the sum over all the scatterers in the medium and all the emitting elements, provided that the round-trip time corresponds to $t_R^j$. 

\bibliographystyle{IEEEtran}
\bibliography{sample} 


\end{document}